\documentclass[a4paper]{article}

\usepackage[numbers]{natbib}
\usepackage{graphicx}
\usepackage{amsmath,slashed}
\usepackage[bookmarksnumbered, pdfauthor={Giovanni Ramirez Garcia
  (ramirez@ecfm.usac.edu.gt)}, pdftitle={Quantum Entanglement In Inhomogeneous
  1D Systems}, pdfsubject={To present a more didactic exposure of basic
  concepts of the rainbow system for the students attending the Latin American
  School of Physics ``Marcos Moshinsky'' 2017.}, pdfkeywords={quantum
  entanglement, rainbow systems}]{hyperref}
\usepackage[affil-it]{authblk}

\title{Quantum Entanglement In Inhomogeneous 1D Systems}
\author{Giovanni Ram\'{\i}rez\thanks{ramirez@ecfm.usac.edu.gt}}
\affil{Instituto de Investigaci\'on, Escuela de Ciencias F\'{\i}sicas y
  Matem\'aticas. Universidad de San Carlos de Guatemala. Guatemala}

\begin{document}
\maketitle

\begin{abstract}
  The entanglement entropy of the ground state of a quantum lattice model with
  local interactions usually satisfies an area law.  However, in 1D systems
  some violations may appear in inhomogeneous systems or in random systems.
  In our inhomogeneous system, the inhomogeneity parameter, $h$, allows us to
  tune different regimes where a volumetric violation of the area law appears.
  We apply the strong disorder renormalization group to describe the maximally
  entangled state of the system in a strong inhomogeneity regime. Moreover, in
  a weak inhomogeneity regime, we use a continuum approximation to describe
  the state as a \emph{thermo-field double} in a conformal field theory with
  an effective temperature which is proportional to the inhomogeneity
  parameter of the system.  The latter description also shows that the
  universal scaling features of this model are captured by a massless Dirac
  fermion in a curved space-time with constant negative curvature $R=−h^2$,
  providing another example of the relation between quantum entanglement and
  space-time geometry.  The results we discuss here were already published
  before, but here we present a more didactic exposure of basic concepts of
  the rainbow system for the students attending the Latin American School of
  Physics ``Marcos Moshinsky'' 2017.
\end{abstract}

\section{INTRODUCTION}
\label{sep:introduction}

For bipartite systems, the von Neumann entropy of the reduced density matrix
is a measure of the quantum entanglement \cite{Amico_etal.RMP.2008}.  The
entanglement entropy of the ground state of quantum lattice 1D models
satisfies an area law which also provides a link between geometry and quantum
structure \cite{Ramirez_etal.JSTAT.2015}.  For the area law in 1D, the
Hastings theorem \cite{Hastings.JSTAT.2007} was proved for local Hamiltonians
with finite interaction and a gap in the energy spectrum.  Moreover,
violations to the area law in 1D may appear in non-local Hamiltonians with
divergent interactions or gapless systems.  In the latter case, for
translational invariant gapless systems which can be described by a conformal
field theory (CFT), the area law is restored by a massive perturbation, thus
the entanglement entropy is proportional to the logarithm of the correlation
length in the scaling regime \cite{Calabrese_Cardy.JSTAT.2004}.

In this work we present a much stronger violation of the area law.  We deform
a 1D critical Hamiltonian with open boundary conditions (OBC) by choosing the
exchange couplings from an exponentially-decay function from the centre of the
chain.  This decrease of the exchange couplings yields to a vanishing gap in
the thermodynamic limit \cite{Ramirez_etal.JSTAT.2014b}.  

We are able to use just one parameter to describe the inhomogeneity of the
system, i.e. the decay rate of the exchange couplings.  In terms of that
inhomogeneity parameter, $h$, we study two different regimes: the strong
disorder regime and the weak disorder regime. An analysis in terms of
entanglement entropy and entanglement spectrum was used to discuss the
intermediate regime where there is a smooth crossover between the critical
system and the maximally entangled ground state without a phase-transition
\cite{Ramirez_etal.JSTAT.2014b}.

\begin{figure}
  \centerline{\includegraphics[width=.85\textwidth]{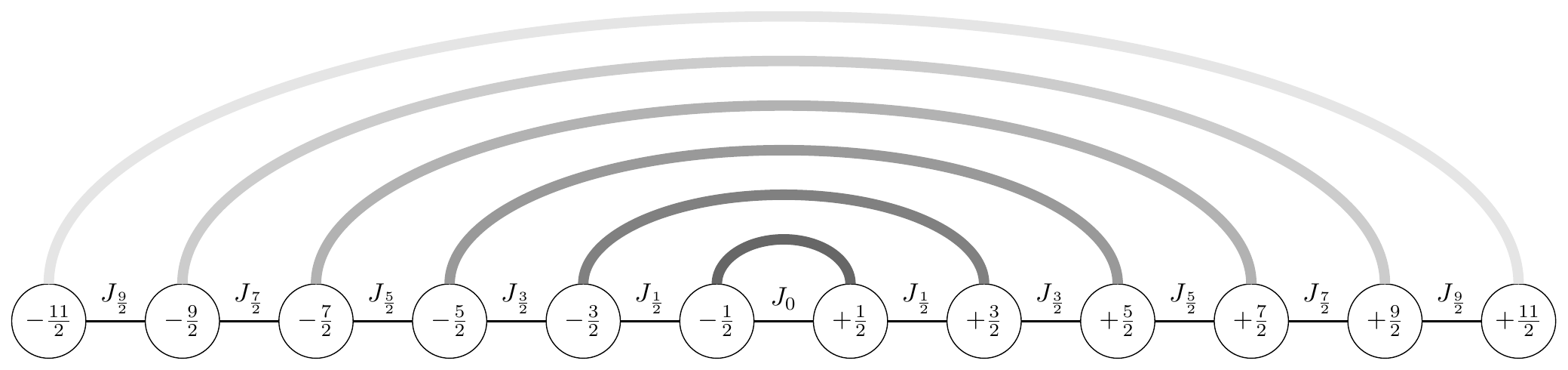}}  
  \caption{Rainbow model using a system of $2L=12$ sites, which are labeled
    with half-integers $\pm m/2$ index in order to simplify the notation.
    Values $\{J_m\}$ represent the power-law decreasing exchange couplings
    between next neighbours.  The single bonds connecting the $(-k/2,k/2)$
    sites represent Bell pairs established at different scales of energy. Bold
    grey colour represents a larger scale of energy.  A full-colour version of
    this diagram can be found in \cite{Ramirez_etal.JSTAT.2014b} where the
    exchange couplings are given in a logarithmic scale.}
  \label{fig:rainbow}
\end{figure}

In the strong disorder regime, i.e. the exchange couplings decay very fast, we
use the Dasgupta-Ma Renormalization Group \cite{Dasgupta_Ma.PRB.1980}, which
was also reformulated to study fermionic systems
\cite{Ramirez_etal.JSTAT.2015}, in order to describe the ground state as a
valence bond state formed by bonds joining the sites located symmetrically
with respect to the centre.  This state was also termed as the
\emph{concentric singlet phase} \cite{Vitagliano_etal.NJP.2010} and has a
rainbow-like structure illustrated in Figure \ref{fig:rainbow}.  In the weak
inhomogeneity regime, we use a continuum approximation to describe the state
as a \emph{thermo-field double} in a CFT with an effective temperature which
is proportional to the inhomogeneity parameter of the system.  This latter
description also shows that the universal scaling features of this model are
captured by a massless Dirac fermion in a curved space-time with constant
negative curvature $R=−h^2$.  Such relation between quantum entanglement and
geometry helps explain thermal effects in quantum field theories studied on
curved backgrounds, e.g. the Unruh effect where from the point of view of an
accelerated observer, the temperature measured in a Minkowski vacuum must be
proportional to its acceleration \cite{RoblesLaguna2017}.

Experiments using ultracold atomic gases \cite{lewenstein2012} have been used
as quantum simulators in order to explore different relations between quantum
mechanics and curved space-time, such as the Dirac equation in an artificial
curved space-time \cite{Boada_etal.NJP.2011}, simulation of $D+1$ dimensions
using a $D$-dimensional quantum system based in optical lattices
\cite{BoadaPRL}, or the dynamics of quantum many-body systems in manifolds
with a non-trivial topology like a M\"obius strip \cite{BoadaNJP}.  Thus, in
this work we present how to relate a tunable parameter, i.e. the inhomogeneity
in exchange couplings, to a curved space-time which may be also seen as a CFT
with an effective temperature.

This work is organised as follows, in a first section we introduce the rainbow
model and present the methods used to obtain the entanglement entropy.  In a
second section we discuss the behaviour of the entanglement entropy as an
evidence of the violation of the area law.  Furthermore, we analyse the system
in both regimes, the strong disorder and the weak disorder, where we present
the quantum state obtained in both regimes: a maximally entangled state and a
thermal state.  After that, we discuss the relation between the inhomogeneous
system and a curved space-time.  In a final section we present some
conclusions and some future works.

\section{INHOMOGENEOUS 1D SYSTEMS}
\label{sec:systems}
We are interested in inhomogeneous systems because, as we will show, they can
present a strong violation of the area law.  The inhomogeneity of the exchange
couplings in many-body systems has been addressed from different points of
view, e.g. smooth changes can be regarded as position-dependent speed of
propagation for the excitation, such as a local gravitational potential
\cite{Boada_etal.NJP.2011} or a horizon \cite{Ramirez_etal.JSTAT.2014b}.
Moreover, an exponential dependence of the exchange couplings with the
position is a characteristic of Kondo-like problems
\cite{Okunishi_Nishino.PRB.2010} and systems with an hyperbolic dependence of
the couplings with the position has been used to study the scaling properties
of non-deformed systems \cite{Ueda_Nishino.JPSJ.2009, Ueda_etal.PTP.2010}.
However, smoothed boundary conditions, i.e. the couplings fall to zero near
the borders, have been used to reduce the finite-size effects when measuring
bulk properties of the ground state \cite{Vekic_White.PRL.1993}.

\subsection{The Rainbow Model}
\label{sec:model}
Consider a fermionic system with $2L$ sites which are labeled with
half-integers $\pm m/2$, in order to simplify the notation.  As usual,
positive numbers are used for sites placed from the centre of the system to
its right border, cf. Figure \ref{fig:rainbow}.  For OBC, the dynamics of the
system is given by the Hamiltonian
\begin{equation}
  \label{eq:rainbow}
  H = -\frac{J_0}{2} c^\dagger_{\frac{1}{2}}c_{-\frac{1}{2}}
  -\sum_{m=\frac{1}{2}}^{L-\frac{3}{2}} \frac{J_m}{2} \left[
    c^\dagger_m c_{m+1} + c^\dagger_{-m} c_{-m-1} \right] +h.c.
\end{equation}
where $J_0$ sets the scale for the exchange couplings, $c_m$ and $c^\dagger_m$
denote the annihilation and creation operators of a spinless fermion at the
site $m$.  The values $\{J_m\}$ follow an exponentially-decay function from
the centre of the chain
\begin{displaymath}
  \label{eq:jotas}
  J_m = J_0 e^{-hm},
\end{displaymath}
$h\geq 0$ characterises the inhomogeneity of the exchange couplings, i.e. it
determines the inhomogeneity of the system.  The case $h=0$ corresponds to the
standard uniform Hamiltonian of a spinless free fermion with OBC
\cite{Ramirez_etal.JSTAT.2014b, Ramirez_etal.JSTAT.2015} and its low energy
properties are captured by a CFT with central charge $c=1$, i.e. the massless
Dirac fermion theory, or equivalently a Luttinger liquid with Luttinger
parameter $K=1$ \cite{Rodriguez_etal.2017}.

It has been shown that other decay functions also results in concentric
singlet phases \cite{Vitagliano_etal.NJP.2010}. Moreover, only an
exponentially-decay function allows us to move smoothly from the homogeneous
case, $h=0$, to a maximally entangled state.

\subsection{Entanglement Over The Rainbow}
\label{sec:entanglement}
In case the exchange couplings have very different values, we can apply the
strong disorder renormalization group (SDRG) method
\cite{Dasgupta_Ma.PRB.1980} which was reformulated to include the fermionic
nature of the particles \cite{Ramirez_etal.JSTAT.2015}.  The RG method starts
by picking up the strongest coupling and then to establish a singlet bond (a
valence bond state) on top of it and then using an effective coupling, which
is obtained using second order perturbation theory, between the two neighbours
of the singlet
\begin{equation}
  \label{eq:rg}
  \tilde J = - \frac{J_L J_R}{J_{max}},
\end{equation}
where $J_L$ and $J_R$ are, respectively, the left and right couplings to the
maximal one, $J_{max}$, which establish the scale of energy where the RG step
is done.  A detailed procedure to obtain the relation in (\ref{eq:rg}) for
spins was presented by Casa Grande et al. \cite{CasaGrande} and the arguments
to extend the method to fermionic particles was presented by Ramirez et
al. \cite{Ramirez_etal.JSTAT.2015}.  See in Figure \ref{fig:rainbow}, the
different scales of energy are represented by the grey scale, a bold colour
represents more energy.

\begin{figure}
  \centerline{\includegraphics[width=.85\textwidth]{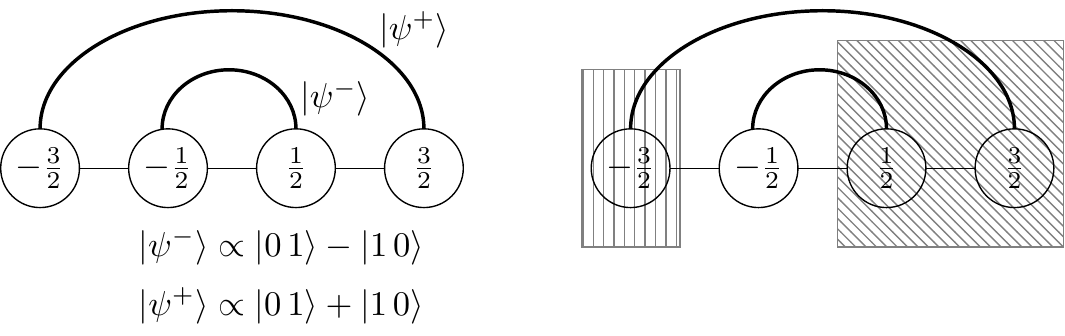}}
  \caption{Left panel: Bell pair states localised at different RG steps. In a
    first RG step, the singlet-type state, $| \psi ^-\rangle$, is established
    with a bond connecting lattice sites $(-1/2,1/2)$. In a second RG step the
    triplet-type state, $| \psi ^+\rangle$, is established with a bond
    connecting lattice sites $(-3/2,3/2)$.  The sign change from negative to
    positive relates the fermion nature of the particles.  Right panel:
    Examples of the blocks of lattice sites which can be obtained in different
    bipartitions of the system.  The block with the vertical-line pattern
    contains only one site, it represents the system's bipartition $[-3/2]\cup
    [-1/2,1/2,3/2]$, and its block-entropy is proportional to the number of
    bonds connecting the block with the rest of the system.  The block with
    the diagonal-line pattern contains two sites representing other
    bipartition, each site with a bond connecting the rest of the system.}
  \label{fig:bonds}
\end{figure}

The renormalization step represents to localise a Bell pair state, or a bond,
between sites connected by $J_{max}$, cf. left panel Figure \ref{fig:bonds}.
Moreover, equation (\ref{eq:rg}) implies that the effective couplings can be
either positive or negative in relation to the nature of the established bond
\cite{Ramirez_etal.JSTAT.2015}, thus there are singlet-type bonds
$|\psi^-\rangle \propto |01\rangle -|10\rangle$ and triplet-type bonds
$|\psi^+\rangle \propto |01\rangle +|10\rangle$.  Both types of bonds share
many properties such as the entanglement \cite{Ramirez_etal.JSTAT.2015}.  The
procedure is repeated until the $2L$ sites are linked by $L$ bonds
establishing a valence bond structure.  See left panel in Figure
\ref{fig:bonds} for a graphical example of two bonds established in successive
RG steps.  The first bond establishes a singlet-type bond $|\psi^-\rangle$
connecting sites $(-1/2,1/2)$ at an energy-scale given by $J_0$.  The
distribution of values of $J_m$ is such that the second RG step establishes a
bond connecting sites $(-3/2,3/2)$, however this is a triplet-type bond.

Once the valence bond structure is completed, the von Neumann entropy for a
block $B$ can be easily computed with
\begin{equation}
  \label{eq:counting}
  S(B) = n_B \log{(2)},
\end{equation}
where $n_B$ is the number of bonds joining $B$ with the rest of the system,
i.e. the number of broken Bell pairs.  See right panel in Figure
\ref{fig:bonds} for a graphical example of two blocks which may be obtained in
different system's bipartitions.  The block with the vertical-line pattern is
obtained with the system's bipartition $[-3/2]\cup[-1/2,1/2,3/2]$ thus, the
block is formed by only one site, i.e. its size is one, which is connected to
the rest of the system by one bond.  The block with diagonal-line pattern is
obtained splitting the system in half, here there are two sites each one with
a bond connecting the block with the rest of the system.  The computation of
the von Neumann entropy becomes a process of counting the bonds connecting the
block of interest.  Nevertheless, in the RG approximation, all the R\'enyi
entropies take the same value of the von Neumann entropy
\cite{Ramirez_etal.JSTAT.2014b}.

Since the Hamiltonian (\ref{eq:rainbow}) is quadratic in the fermionic
operators, the exact diagonalization method \cite{Ramirez_etal.JSTAT.2014b}
can be used to obtain the properties of the system using a canonical
transformation for the fermionic operators $c^\dagger$ and $c$, such that
\begin{displaymath}
  \label{eq:canonicalTransf}
  \bar{\psi}^k = \sum_i v_{k,i} c_i^\dagger,
\end{displaymath}
are the single-body modes, with single-body energy $\epsilon_k$ associated,
for which the Hamiltonian (\ref{eq:rainbow}) is diagonal, $v_{k,i}$ is a
unitary matrix obtained by the diagonalization of the hopping matrix.  Thus,
the ground state of the system is obtained by filling $L$ fermions in the set
$\Omega$ of single-body modes with energy below the Fermi level, i.e.
$\Omega=\{k|\epsilon_k<\epsilon_F\}$.  These occupied single-body modes,
$\psi^k$, can be used to compute the correlation matrix
\cite{Ramirez_etal.JSTAT.2014b, Ramirez_etal.JSTAT.2015}
\begin{displaymath}
  \label{eq:corrMat}
  C_{ij} = \langle GS|c^\dagger_i c_j|GS\rangle = \sum_{k\in \Omega} \bar
  \psi^k_i\psi^k_j,
\end{displaymath}
where $i$ and $j$ label sites which are both inside the considered block $B$.
The eigenvalues $\{\nu_p\}$ of the correlation matrix determine the $n$-order
R\'enyi entropies for the block $B$ of the system, so
\begin{displaymath}
  \label{eq:renyi}
  S^{(n)} (B) = \frac{1}{1-n} \sum_p \log{[\nu_p^n +(1-\nu_p)^n]},
\end{displaymath}
and the von Neumann entropy can be obtained from the limit $n\to 1$.  The
scaling behaviour of the entanglement entropy of a block $B$ of size $\ell$ is
known to be
\begin{equation}
  \label{eq:entropyScaling}
  \begin{split}
    S^{(n)}(\ell) \simeq &\frac{c}{12} \left( 1+\frac{1}{n}\right) \log{\left[
        \frac{4L}{\pi} \sin{\left( \frac{\pi \ell}{2L}\right)} \right]} +c'_n \\
    &+f_n \cos{(\pi\ell)}\left[ \frac{8L}{\pi} \sin{\left(
          \frac{\pi\ell}{2L}\right)}\right]^{-K/n},
  \end{split}
\end{equation}
where the first term is given by the CFT associated for homogeneous free
fermion model, $c'_n$ is a non-universal constant, the last term accounts for
the fluctuations at Fermi momentum, $k_F=\pi/2$, and the constants $f_n$ can
be obtained analytically \cite{Fagotti_Calabrese.JSTAT.2011}
\begin{displaymath}
  \label{eq:falpha}
  f_n = \frac{2}{1-n}\left[ \frac{\Gamma\left(
        \frac{1}{2}+\frac{1}{2n}\right)}{\Gamma\left(
        \frac{1}{2}-\frac{1}{2n}\right)} \right],
\end{displaymath}
where $f_1\equiv 1$ by definition.

\section{RESULTS}
\label{sec:results}
Using the methods described above, we are able to obtain the entanglement
properties of the system.  Selecting the inhomogeneity parameter, $h$, we are
able to set the system in two regimes: The strong disorder and the weak
disorder regime.  The RG method explains with simple terms the violation to
the area law, but it describes all the R\'enyi entropies to be the same as the
von Neumann entanglement entropy.  Nevertheless, the exact diagonalization
method is applicable to all values of $h$.  It was showed that the system
moves in the two regimes without phase transitions
\cite{Ramirez_etal.JSTAT.2014b}.

To give a full description of the rainbow model first we show that the
violation of the area law appear for $h\geq 0$.  Although for $h=0$, the
logarithmic violation is well described by CFT, there appears a volumetric
behaviour in the entanglement entropy for $h>0$.  Thus, we apply the RG method
for $h\gg 0$, i.e. the strong disorder regime, to describe the structure of
the ground state in terms of the valence bond structure obtained in each RG's
step.  Furthermore, we use an analytic continuation in the limit $h\ll 1$,
i.e. the weak disorder regime, to describe the ground state as a thermal
state.  Moreover, we show that this state is related to the one described by a
massless Dirac fermion theory in a curved space-time, thus this analysis could
help to explain thermal effects in quantum field theories studied on curved
backgrounds.

\subsection{Violation of the Area Law}
\label{sec:violation}
\begin{figure}
  \centerline{\includegraphics[width=.5\textwidth]{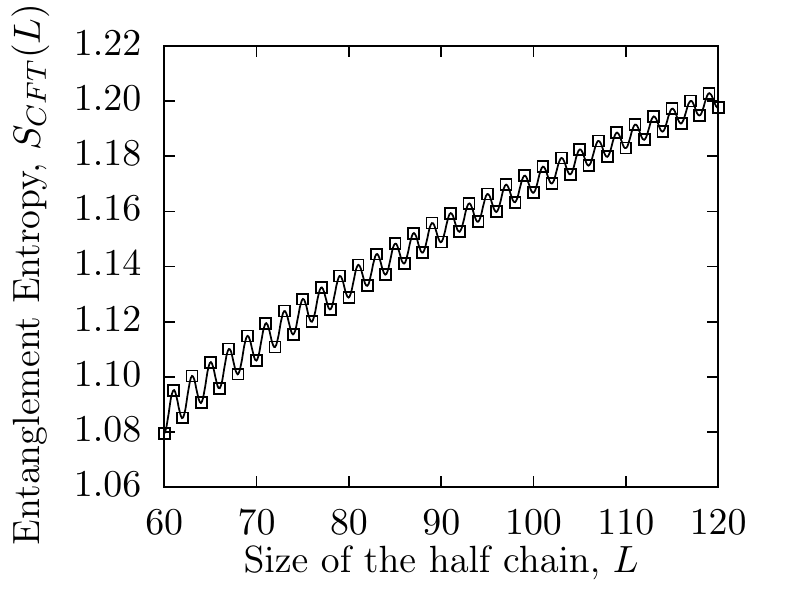}
    \includegraphics[width=.5\textwidth]{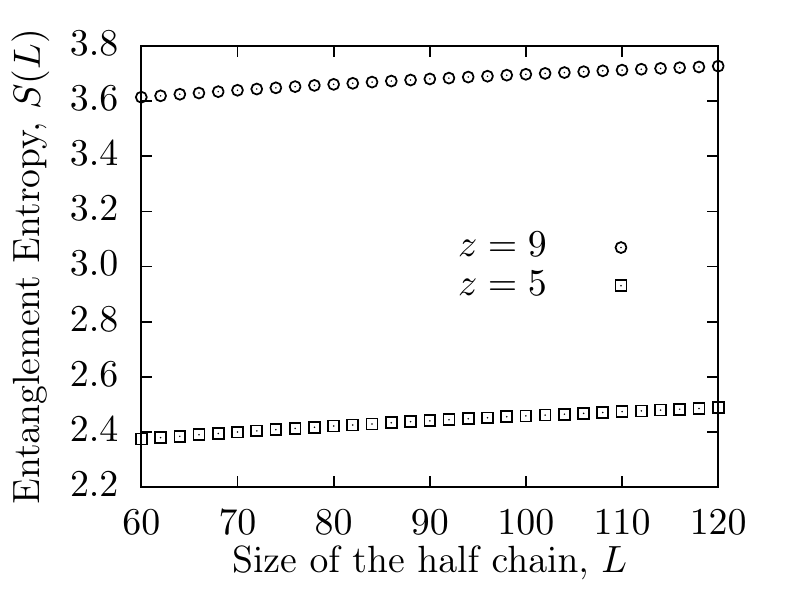}}
  \caption{Left: Entanglement entropy for one half of the chain $S_{CFT}(L)$
    as a function of the half size of the system $L$ for the uniform
    system. The line corresponds to the fit given to Equation
    (\ref{eq:entropyCFT}).  Right: Entanglement entropy for the inhomogeneous
    system as a function of the half size of the system for different values
    of $z\equiv hL$.}
  \label{fig:blockEntropy}
\end{figure}

As stated above, the area law is violated for all the values of the
inhomogeneity parameter.  The logarithmic violation appear in the limit $h=0$,
i.e. the uniform system, it corresponds to a spinless free fermion system, its
low energy properties are described by a CFT with $c=1$.  The entanglement
entropy for a block of size $\ell$ is given in equation
(\ref{eq:entropyScaling}).  However, we focus in blocks of one half of the
system in order to simplify the analysis of the functional dependence of the
entropy, thus, the entanglement entropy is \cite{Ramirez_etal.JSTAT.2014b}
\begin{equation}
  \label{eq:entropyCFT}
  S_{CFT} (L) = \frac{c}{6} \log{(L)} + c' +f_1 \frac{\cos{(\pi L)}}{L^{K}},
\end{equation}
where $c'$ is an additive constant which includes the boundary entropy and
non-universal contributions \cite{Calabrese_Cardy.JSTAT.2004, Vidal_etal.03},
$f_1$ is a measure of the amplitude of the oscillations, which were shown to
vanish as the inhomogeneity of the system increases
\cite{Ramirez_etal.JSTAT.2014b}.  The left panel in Figure
\ref{fig:blockEntropy} shows a fit of the half chain entanglement entropy,
Equation (\ref{eq:entropyCFT}), for different values of the system half chain
$L$.  This logarithmic behaviour represents a violation of the area law with a
leading term proportional to $\log{(L)}$.

The volumetric violation of the area law appear for all values $h>0$.  It is
convenient to define an \emph{effective} size of the system $z\equiv hL$ for
the given inhomogeneity in the system in order to simplify the functional
analysis and the notation.  In terms of this effective size of the system, the
strong disorder regime appear for $z\gg 1$ and the weak disorder regime appear
for $z\ll 1$.  This change in the effective size of the system anticipates
some relation between the inhomogeneity of the system and its geometry.

The right panel in Figure \ref{fig:blockEntropy} shows the entanglement
entropy of the half of the chain as a function of the system size for $z=5$
and for $z=9$, i.e. both $L$ and $h$ varies in order to keep $z$ constant.
The entanglement entropy shows a linear behaviour which was explained as
\cite{Ramirez_etal.JSTAT.2014b}
\begin{equation}
  \label{eq:EntropyGEN}
  S(L) = \frac{c(z)}{6} \log{(L)} + d(z) + f(z) \frac{\cos{(\pi L)}}{L^K},
\end{equation}
where $c(z)$, $d(z)$ and $f(z)$ are now functions of $z$ and explain the
behaviour of the entanglement entropy for different values of the
inhomogeneity.  For the limit $z\gg 1$, functions $c(z)\to 0$, $d(z)\to
0.318z$ and $f(z)\to 0$ \cite{Ramirez_etal.JSTAT.2014b} thus, the entanglement
entropy is expected to have a linear behaviour $S(L) \to 0.318\, h L $, but in
the strong inhomogeneity limit it should behave as $S(L)=L\log{(2)}$ as
discussed above.

\begin{table}
  \caption{Summary of the different Area law's violation for all the regimes
    of the Rainbow Model \label{tab:area}}
\begin{tabular}{lccl}
\hline
& & \textbf{Violation}  & \textbf{Leading} \\
\textbf{Regime} & \textbf{Description} & \textbf{type} & \textbf{term} \\
\hline
$h=0$    (uniform)  & CFT & logarithmic    & $\propto \log{(L)}$\\
$h\ll 1$ (weak disorder) & thermal-state & volumetric  & $\propto L$\\
$h\gg 1$ (strong disorder) & valence bond states & volumetric  &  $\propto L$\\
\hline
\end{tabular}
\end{table}

In Table \ref{tab:area} we present a summary of the violations of the area law
for the different regimes of the rainbow model.  Note that the area law is
violated in all cases by the model.  The model can be reduced to its
homogeneous, or uniform, case for $h=0$ where the CFT associated characterises
a ground state which violates the area law with a logarithmic dependence of
the size of the system.  The cases of weak disorder and strong disorder have a
ground state which violates the area law with a volumetric behaviour because
the function $d(z)$, cf. Equation (\ref{eq:EntropyGEN}), becomes the leading
term as the inhomogeneity of the system grows. It should be noticed that the
function $c(z)$ decreases monotonously as the inhomogeneity of the system
grows in agreement to the Zamolodchikov c-theorem
\cite{Ramirez_etal.JSTAT.2014b}.

\subsection{Strong Disorder Regime}
\label{sec:strong}
This regime is well explained by the SDRG described above.  At the beginning of
the RG method, the central link is used to renormalize the two central sites,
it establishes the scale of energy.  The effective link, obtained with the
Equation (\ref{eq:rg}), sets the new scale of energy, and the procedure is
repeated. When the RG procedure is done, the GS can be obtained as the product
of valence bond states connecting the $(-k/2,k/2)$ sites.  Nevertheless, in
terms of the bonding and anti-bonding operators defined as
\cite{Ramirez_etal.JSTAT.2015}
\begin{eqnarray}
  \label{eq:bonding}\nonumber
  (b^+_{ij})^\dagger &=& \frac{1}{\sqrt{2}}
  \left(c^\dagger_i+c^\dagger_j\right), \\\label{eq:antibonding} \nonumber
  (b^-_{ij})^\dagger &=& \frac{1}{\sqrt{2}}
  \left(c^\dagger_i-c^\dagger_j\right), 
\end{eqnarray}
the GS of the Hamiltonian (\ref{eq:rainbow}) is given by
\begin{equation}
  \label{eq:RB}
  |R(L)\rangle \equiv 
  \left( b^{S_L}_{-L+\frac{1}{2},L-\frac{1}{2}} \right)^\dagger \cdots
  \left( b^{+}_{-\frac{5}{2},\frac{5}{2}} \right)^\dagger 
  \left( b^{-}_{-\frac{3}{2},\frac{3}{2}} \right)^\dagger 
  \left( b^{+}_{-\frac{1}{2},\frac{1}{2}} \right)^\dagger |0\rangle,
\end{equation}
where $S_L=(-1)^L$.  The entanglement entropy for a block is obtained by
counting the number of links connecting the block with the rest of the system,
cf. Figure \ref{fig:bonds}.  For a block of the half of the system, $n_B=L$
thus, we obtain a maximally entangled state and it fulfills a volume law.  The
energy gap can be estimated as the effective energy of the last bond
established, i.e. $\alpha^{2L}$ in Figure \ref{fig:rainbow}.  Thus the gap
vanishes in the limit $L\to\infty$ in agreement with the Hastings theorem
\cite{Ramirez_etal.JSTAT.2014b}.

\subsection{Weak Disorder Regime}
\label{sec:weak}
In this regime, a continuum approximation of the Hamiltonian
(\ref{eq:rainbow}) is obtained by expanding the local operator $c_n$ into the
slow chiral modes, $\psi_R(x)$ and $\psi_L(x)$ around the Fermi points $\pm
k_F$
\begin{equation}
  \label{eq:continuum}
  \frac{c_n}{\sqrt{a}} \simeq e^{ik_F x} \psi_L(x) + e^{-ik_F x} \psi_R(x),
\end{equation}
and located at the position $x=an \in (-{\cal L, L})$, where $a={\cal L}/L$ is
the lattice spacing.  Using Equation (\ref{eq:continuum}) in
(\ref{eq:rainbow}) we obtain
\begin{equation}
  \label{eq:contHam}
  H \simeq \frac{ia}{2} \int_{-\cal L}^{\cal L} dx\; e^{-\frac{h|x|}{a}} \left[
    \psi^\dagger_R \partial_x \psi_R -(\partial_x\psi^\dagger_R)\psi_R
    -\psi^\dagger_L\partial_x\psi_L +(\partial_x\psi^\dagger_L)\psi_L \right],
\end{equation}
assuming that the fields $\psi_{R,L}(x)$ vary slowly with $x$ in order to drop
some cross terms.  This Hamiltonian (\ref{eq:contHam}) describes the low
energy excitations of the original lattice Hamiltonian and can also be brought
to the standard canonical form of a free fermion with OBC
\cite{Ramirez_etal.JSTAT.2015}.  A change in the variables
\begin{equation}
  \label{eq:trafo}
  \tilde x = \mathrm{sign}(x) \frac{e^{h|x|}-1}{h},
\end{equation}
in terms of the inhomogeneity of the system $h$, maps the interval $x\in
[-L,L]$ into the interval $\tilde x \in [-\tilde L, \tilde L]$.  The
transformation for the fermion fields are
\begin{displaymath}
  \label{eq:fermionTilde}
  \tilde \psi_{R,L} (\tilde x) = \left( \frac{d\tilde x}{dx}\right)^{-1/2}
  \psi_{R,L} (x),
\end{displaymath}
which can be used to transform the Hamiltonian (\ref{eq:contHam}) to 
\begin{displaymath}
  \label{eq:tildeHam}
  H \simeq i \int_{-\tilde L}^{\tilde L} d\tilde x \left[ \tilde\psi_R^\dagger
    \partial_{\tilde x} \tilde\psi_R -\tilde\psi_L^\dagger\partial_{\tilde
      x}\tilde\psi_L\right],
\end{displaymath}
which is the free fermion Hamiltonian for a chain of size $2\tilde L$.  A
prediction of the entanglement entropy for the deformed chain is given by
\cite{Ramirez_etal.JSTAT.2015}
\begin{displaymath}
  \label{eq:Scsp}\displaystyle
  S(L) = \frac{c}{6} \log{\left( \frac{e^{hL}-1}{h}\right)} +c'
  = \frac{c}{6} \log{(\tilde L)} + c',
\end{displaymath}
for larger systems, which can be compared with the equation given by the CFT
in Equation (\ref{eq:entropyCFT}) by means of the transformation of variables
given in Equation (\ref{eq:trafo}).  This relation allowed the comparison of
the entanglement entropy in the limit $h\to 0^+$ with the entropy of a thermal
state at temperature $T=1/\beta$ in a CFT \cite{Calabrese_Cardy.JSTAT.2004} in
order to relate
\begin{equation}
  \label{eq:temp}
  T = \frac{1}{\beta} = \frac{h}{2\pi},
\end{equation}
thus, the rainbow state is similar to a thermal state with an effective
temperature proportional to the inhomogeneity of the system, $h$.

Recently, the relation between quantum entanglement and the geometry of the
space-time where the system lives motivates the study of links between
geometry and quantum structure given the area-laws.  It was shown that the
rainbow Hamiltonian can be obtained from the action of a massless Dirac
fermion in a curved space-time \cite{Rodriguez_etal.2017}.  

The interest in models with smoothly varying couplings and quantum field
theories in curved space-times has been put forward in
\cite{Boada_etal.NJP.2011, RodriguezLagunaPRA}, with potential cold atom
realisations in sight. In addition to those proposals for tabletop experiments
that mimic effects from high-energy physics, there is another good reason for
exploring the relevance of quantum field theory in curved space for the
physics of ultracold quantum gases. Most experimental setups involve trapping
potentials, often harmonic ones, that result in inhomogeneous density profiles
in the models one wants to simulate; in fact, the presence of non-uniform
potential wells, or background electromagnetic fields, is the rule rather than
the exception. It was pointed out very recently \cite{dubail2017}, in the
study of non-interacting fermion gases in $1+1$ dimensions, that the
inhomogeneity generated by the trapping potential are captured by quantum
field theory in curved space-time.

This derivation is done using the covariant formalism of relativistic field
theory with Minkowski signature.  The Lagrangian associated to the Hamiltonian
(\ref{eq:rainbow}) is
\begin{equation}
  \label{eq:lagrangian}
  \begin{split}
    {\cal L} = &\psi_-^\dagger\partial_0\psi_- + \psi_+^\dagger \partial_0
    \psi_+ + e^{-h|x^1|} \left( \psi_-^\dagger\partial_1\psi_-
      -\psi_+^\dagger\partial_1 \psi_+\right) \\&+ \frac{h}{2} \mathrm{sign}(x^1)
    e^{-h|x^1|} (\psi_+^\dagger\psi_+ -\psi_-^\dagger\psi_-),
  \end{split}
\end{equation}
where $\psi_- = \psi_L$, $\psi_+=\psi_R$, and the coordinates are $(t,x) =
(x^0,x^1)$ and $\partial_\mu=\partial/\partial x^\mu$.  For $h=0$ the Equation
(\ref{eq:lagrangian}) becomes the Dirac action of a massless fermion in $1+1$
dimensions.  The flat space-time metric, denoted by $\eta^{ab} (a,b=0,1)$, has
the signature $(-,+)$. In order to express the Lagrangian
(\ref{eq:lagrangian}) as the Dirac action in curved space-time we need the
metric $g_{\mu\nu}$, the zweibein $e^a_\mu$, its inverse $E^\mu_a$, the spin
connection $\omega_\mu^{ab}$ and the Christoffel symbols
$\Gamma^\nu_{\lambda\mu}$ which are related as
\begin{eqnarray}
  \nonumber
  g_{\mu\nu} &=& \eta_{ab} e^a_\mu e^b_\nu,\\
  \nonumber\label{eq:relations}
  E^\mu_a &=& g^{\mu\nu} \eta_{ab} e^b_\nu,\\
  \nonumber
  \omega_\mu^{ab} &=& e^a_\nu \partial_\mu E^{b\nu} + e^a_\nu E^{b\lambda}
  \Gamma^\nu_{\lambda\mu},
\end{eqnarray}
now, the Dirac Lagrangian of a massless fermion in curved space-time is
\begin{displaymath}
  \label{eq:masslessfermion}
  {\cal L} = e \bar\psi {\slashed D} \psi,
\end{displaymath}
where ${\slashed D}=E^\mu_a\gamma^aD_\mu$ and $e=\mathrm{det}e^a_\mu$, thus
comparing equations and solving for the space-time metric, we obtain
$g_{00}=-e^{-2h|x|}$ and $g_{11}=1$ and with the non-vanishing components of
the Christoffel symbols and the Ricci tensor yield the scalar curvature
\begin{displaymath}
  \label{eq:curvature}
  R(x) = g^{\mu\nu}R_{\mu\nu}(x) = 4h\delta(x) - h^2,
\end{displaymath}
that is constant and negative here except at the origin where it is singular
\cite{Rodriguez_etal.2017}.

\section{CONCLUSIONS}
\label{sec:conclusions}
In this work we presented how to parametrise the inhomogeneity of a 1D system
whose dynamics is described by the local Hamiltonian (\ref{eq:rainbow}).  The
variation of the inhomogeneity parameter, $h$, allowed us to move the system
between two different violations of the area law without a phase transition.

In the strong disorder regime the behaviour of the entanglement entropy is
explained with the structure of the ground state (\ref{eq:RB}). The
entanglement entropy for this state is obtained as described by Equation
(\ref{eq:counting}), counting the number of links connecting the block of
interest with the rest of the system.  It represents a linear growth of the
entanglement entropy with the size of the system. But, since the gap of the
system vanishes in the limit $L\to\infty$ the Hastings theorem is kept.

In the weak inhomogeneity limit, a continuum approximation can be found to
relate the system to a homogeneous one.  The rainbow system can be viewed as a
conformal system at a finite temperature proportional to the inhomogeneity of
the system as can be seen in Equation (\ref{eq:temp}).  The universal scaling
features of the rainbow model are captured by a massless Dirac fermion in a
curved space-time with constant negative curvature.

\section{ACKNOWLEDGEMENTS}
We would like to acknowledge G Sierra and J Rodr\'{\i}guez-Laguna for their
useful comments and support.  We also acknowledge financial support from the
\emph{Direcci\'on General de Docencia} at the \emph{Universidad de San Carlos
  de Guatemala} and from the Organizing Committee of the Latin American School
of Physics ``Marcos Moshinsky'' 2017.

\bibliographystyle{unsrt}
\bibliography{references}
\end{document}